# Traceable precision generation and measurement of pA direct currents

Hansjörg Scherer[*], Gerd-Dietmar Willenberg[*], Dietmar Drung[†], Martin Götz[*], and Eckart Pesel[*]

[*]Physikalisch-Technische Bundesanstalt, Bundesallee 100, 38116 Braunschweig, Germany
hansjoerg.scherer@ptb.de
[†]Physikalisch-Technische Bundesanstalt, Abbestraße 2-12, 10587 Berlin-Charlottenburg, Germany

*Abstract* — We present the comparison of results from the generation and the measurement of direct currents in the pA range, performed with traceable state-of-the-art methods at highest accuracy. The currents were generated with the capacitor charging method, and measured with the first prototype of a novel picoammeter developed at PTB. The results confirm the agreement between both methods and the linearity of the new picoammeter at an uncertainty level of few µA/A. Also, they shed light on the ac-dc difference of capacitors, crucial for current generation with the capacitor charging method.

## I. INTRODUCTION

Small electrical currents are of increasing interest in the fields of fundamental and practical metrology. Recent advances in the field of single-electron transport devices offer ways for generating currents of the order of 100 pA with uncertainties of one part per million [1] or better [2] − [4]. This is expected to have fundamental impact on the future quantum-based realization of the SI unit ampere [5]. Regarding the measurement of sub-nA currents there is increasing need for the calibration of picoammeters, e.g. for applications in dosimetry. In order to underpin their calibration and measurement capabilities in this field, lately thirteen National Measurement Institutes (NMIs) have evaluated their corresponding calibration methods by an international comparison for the first time [6].

In this paper we present the results of a comparison of complementary state-of-the-art techniques for the traceable generation and measurement of direct currents in the pA range at highest accuracy. The capacitor charging method was used for the current generation. The measurement of this current was performed using the first prototype of a novel picoammeter, presently under development at PTB.

## II. BACKGROUND AND METHODS

The capacitor charging method [7] is a well-established technique for the generation of small direct currents used by NMIs for standard calibrations [6]. With this method a current is generated by charging or discharging a high-precision, usually gas-filled, capacitor of capacitance $C$ by applying a voltage ramp, i.e. a voltage $V$ linearly increasing or decreasing with slope $dV/dt$. The resulting current $I = C \cdot dV/dt$ is, thus, traced back to the SI units volt, second and farad. The frequency dependence of the capacitance $C$ was found crucial for the accuracy of the method [8]. At best, 100 pA currents can be generated with relative uncertainties of ≈ 10 µA/A [9].

The precision of sub-nA current measurements with conventional commercial picoammeters, even following calibration, is limited to about 10 µA/A due to amplifier gain drifts. This is insufficient for future fundamental and practical metrological applications. Up to now, only one rather complicated setup was reported that achieved 1 µA/A accuracy [1]. To improve this situation, PTB is pursuing the development of a new picoammeter: the "Ultrastable Low-noise Current Amplifier" (ULCA) is a fully non-cryogenic single-box instrument based on specially designed op-amps and resistor networks, working as current-to-voltage converter [10], [11]. For the experiments reported here, a first prototype with an effective transresistance $A_{tr} = U_{out} / I_{in}$ of 50 MΩ with a temperature coefficient of 1.8 µΩ/Ω per Kelvin was used. The current noise level was about 4 fA/√Hz, constant in the frequency range between 0.01 Hz and 0.3 Hz. Full traceability of the measurements was provided by calibrating the amplifier against the quantum Hall resistance (QHR) with a cryogenic current comparator (CCC).

## III. EXPERIMENTAL DETAILS

### A. Current generation with the capacitor charging method

A digital voltage ramp generator was used, developed for picoammeter calibrations at PTB [12]. The computer-controlled instrument is capable of generating highly linear voltage ramps running between -10 V and +10 V with highly stable and linear slope, adjustable between 1 mV/s and 1 V/s. For monitoring the ramp generator output, the slope was measured using an 8 ½ digit multimeter (DMM), triggered by a stabilized precision time base. The ramp generator output was connected to commercial capacitors of type GR1404 (two nominally identical specimens "A" and "B" were used) with nominal capacitances $C$ = 1 nF. They were calibrated by using a commercial precision capacitance ac bridge at the operation frequency of 1 kHz. Current generation was performed in subsequent cycles, each of them consisting of the four phases i) zero, ii) + $I$, iii) zero, and iv) - $I$. The accuracy of the generated currents, ranging from 50 pA to 500 pA by variation of the voltage slopes, corresponded to total relative uncertainties of about 11 µA/A dominated by the contribution from the ac-dc differences of the capacitors.

## B. Current measurement with the ULCA prototype

For the comparison experiments, the ULCA input was connected to the current generator output (i.e. a terminal of the 1 nF capacitor) via a low noise coaxial cable. The temperature was monitored during the measurement for corrections due to the temperature coefficient of $A_{tr}$. The voltage output $U_{out}$ was sampled by an 8 ½ digit DMM which was calibrated against a Josephson voltage standard system. Data acquisition for each current measurement was performed over periods between 8 h and 66 h duration, after which the statistical uncertainty contributions typically were at a level of 1 part per million or below.

Following this, CCC-based calibrations of the ULCA were performed with a current of about 3 nA, tracing the transresistance $A_{tr}$ to the QHR with a total relative uncertainty of 1 μΩ/Ω. The long-term stability of $A_{tr}$ was determined to be better than 10 μΩ/Ω per year.

## IV. RESULTS AND CONCLUSION

The main experimental results are summarized in Fig. 1.

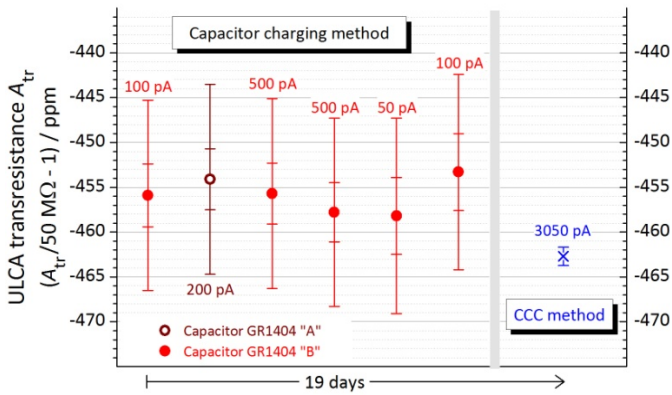

Fig. 1. Comparison of the two independent ULCA calibration methods via capacitor charging (left) and CCC (right panel). All error bars corresponding to standard uncertainties ($k = 1$). For the capacitor charging method, uncertainty bars are shown without (small) and including (large) the contribution due to frequency dependencies of the capacitors used.

The comparison of the two picoammeter calibration methods presented in this paper confirms their agreement within the standard uncertainty ($\approx 11$ μA/A) of the currents generated with the capacitor charging method. Also, the results confirm the linearity of the ULCA picoammeter over a large current range. The transresistance values determined with the capacitor charging method appear slightly ($\approx 5$ to 10 μΩ/Ω), but systematically higher than the value determined from the CCC calibration, which is attributed to uncertainties arising from non-idealities (frequency dependence) of the two capacitors used with the capacitor charging current generation method.

More details and results of ongoing experiments with a further improved ULCA prototype with a higher transresistance ($A_{tr} = 1$ GΩ) will be presented at the conference.

ACKNOWLEDGEMENT

We thank Ralf Behr for the voltmeter calibration. This work was done within Joint Research Project "Qu-Ampere" (SIB07) supported by the European Metrology Research Programme (EMRP). The EMRP is jointly funded by the EMRP participating countries within EURAMET and the European Union.